\documentclass[epj]{svjour}

\usepackage{latexsym}
\usepackage{graphicx}
\usepackage{fancyhdr}
\usepackage{amssymb}
\def\makeheadbox{{
 }}
\def\pplogo{\vbox{\kern-\headheight\kern -29pt
\halign{##&##\hfil\cr&{\ppnumber}\cr\rule{0pt}{2.5ex}&\ppdate\cr}}} \makeatletter
\def\ps@firstpage{\ps@empty \def\@oddhead{\hss\pplogo}
  \let\@evenhead\@oddhead
}

\def\mytitle{My title}
\def\myauthors{My name}
\def\mytype{My type of session}
\def\mysession{My session}

\def\beq{\begin{equation}}
\def\eeq{\end{equation}}
\def\eq{\end{equation}}
\def\ba{\begin{eqnarray}}
\def\ea{\end{eqnarray}}
\def\Qbar{\overline{Q}}
\def\Pbar{\overline{P}}

\def\mytitle{DSB near Enhanced Symmetry Points} 
\def\myauthors{Rouven Essig, Kuver Sinha, and Gonzalo Torroba}    
\def\mytype{Contributed Talk}
\def\mysession{Theoretical Models}

\begin{document}
\title{Enhanced symmetry points and metastable supersymmetry breaking along pseudo-runaway directions}
\author{Rouven Essig\thanks{\emph{Speaker at SUSY07, Karlsruhe, Germany}}, 
 Kuver Sinha, 
 Gonzalo Torroba
}                     
%
%
\institute{NHETC, Department of Physics and Astronomy,
Rutgers University, Piscataway, NJ 08854, U.S.A.
}
%
\date{}

\abstract{
We construct a model with long-lived metastable vacua in which all the
relevant parameters, including the supersymmetry breaking scale, are
generated dynamically by dimensional transmutation. 
The metastable vacua appear along a pseudo-runaway direction near a point
of enhanced symmetry as a result of a balance between non-perturbative
and perturbative quantum effects.  
We show that metastable supersymmetry breaking is a rather generic 
feature near certain enhanced symmetry points of gauge theory moduli 
spaces.
\PACS{
      {11.30.Pb}{Supersymmetry}   \and
      {11.30.Qc}{Spontaneous and radiative symmetry breaking}
     } 
} 
\maketitle
%

\section{Introduction}
\label{intro}

Intriligator, Seiberg and Shih (ISS) \cite{Intriligator:2006dd}
showed that nonsupersymmetric vacua are rather generic if one requires them to
be only local, rather than global, minima of the potential.
One lesson from ISS is that certain properties of moduli spaces can hint at the existence of
metastable vacua.  In their case, it was the existence of supersymmetric vacua coming in from infinity
that signaled an approximate R-symmetry.  Here we will point out that one should also look for another
feature, namely, enhanced symmetry points, which are defined by the appearance of massless particles.
We claim that if the moduli space has certain coincident enhanced symmetry points, metastable
vacua with all the relevant couplings arising by dimensional transmutation may be obtained
\cite{Essig:2007xk}.

The model considered here consists of two SQCD sectors, each with independent
rank and number of flavors, coupled by a singlet.  It involves
only marginal operators with all scales generated dynamically.
At the origin of moduli space, the singlet vanishes and the quarks of
both sectors become massless simultaneously.
There are thus two coincident enhanced symmetry points at the origin.
While one of the SQCD sectors is in the electric range and has a 
runaway, the other has a magnetic dual description as
an O'Raifeartaigh-like model.  Near the enhanced symmetry point, the
Coleman-Weinberg corrections stabilize the nonperturbative instability producing
a long-lived metastable vacuum.
We refer to a runaway direction stabilized by perturbative quantum corrections as a 
``pseudo-runaway''.
A feature of our model is that it may be possible to gauge parts of
its large global symmetry to obtain renormalizable, natural models of direct gauge mediated
supersymmetry breaking with a singlet. R-symmetry is broken both
spontaneously and explicitly in our model; the spontaneous breaking allows for non-zero
gaugino masses, while the explicit breaking allows for a non-zero $R$-axion mass.

\section{Models with Moduli Dependent Masses}
\label{sec:1}

The matter content of the models considered here
consists of two copies of
supersymmetric QCD, each with independent rank and
number of flavors, and a single gauge singlet chiral superfield:
\begin{equation}\label{Eqn:fieldcontent}
{
\matrix{
\quad     & SU(N_c) & SU(N_c^{\prime})   &   \quad      \cr
          &         &             &  \quad       \cr
Q_i         & \square   &    1       & i=1,\dots,N_f      \cr
\Qbar_i     & \overline{\square} &  1  & \quad         \cr
P_{i^{\prime}}        &  1 &    \square         &
    i^{\prime}=1,\dots,N_f^{\prime}      \cr
\Pbar_{i^{\prime}}      & 1 &  \overline{\square} & \quad         \cr
\Phi      &    1    &    1        &   \quad      \cr}}
\end{equation}
The most general tree-level superpotential with only relevant or
marginal terms in four dimensions
for the matter content (\ref{Eqn:fieldcontent}) with $N_c$, $N_c'$ $\ge$ 4
is
\beq\label{Eqn:Wtreegeneral}
W = (\lambda_{ij}  \Phi + \xi_{ij})   Q_i \Qbar_j +
  ( \lambda^{\prime}_{i^{\prime}j^{\prime}} \Phi
  + \xi^{\prime}_{i^{\prime}j^{\prime}})
    P_{i^{\prime}} \Pbar_{j^{\prime}} + w(\Phi) \, ,
\eeq
where $w(\Phi)$ is a cubic polynomial in $\Phi$.
We shall find metastable vacua even in the simplest case of $w(\Phi)=0$, which we assume from now on.
The general situation is discussed in Sec.~4 (in \cite{Brummer:2007ns},
the case $w(\Phi)=\kappa\Phi^3$ was used to stabilize $\Phi$ supersymmetrically).

At the classical level this superpotential has an accidental
$U(1)_R \times U(1)_V \times U(1)_V^{\prime}$ global symmetry.
In the quantum theory the $U(1)_R$ symmetry is anomalous
with respect to the  $SU(N_c)$ and $SU(N_c^{\prime})$ gauge dynamics.
In the $SU(N_f)_V \times SU(N_f^{\prime})_V$ global symmetry
limit the superpotential (\ref{Eqn:Wtreegeneral}) (with $w(\Phi)=0$) reduces to
\beq\label{Eqn:Wtreem}
W = (\lambda \Phi + \xi) {\rm tr}(   Q \Qbar) +
  (\lambda^{\prime} \Phi + \xi^{\prime})  {\rm tr}
    (P \Pbar )\,.
\eeq
For $\xi=\xi^{\prime}$ both masses may simultaneously
be absorbed into a shift of $\Phi$, and the tree level superpotential
in this case reduces to
\beq\label{Eqn:Wtree}
W = \lambda \Phi  ~{\rm tr}(   Q \Qbar) +
  \lambda^{\prime} \Phi~   {\rm tr}
    (P \Pbar )\,.
\eeq

The classical moduli space of vacua is as usual parameterized by baryon 
and meson invariants, see \cite{Essig:2007xk}.
It is lifted by nonperturbative effects in the quantum theory.
Since the metastable supersymmetry breaking
vacua discussed below arise for
$\Phi \neq 0$, only this branch of the moduli space will be considered
in detail.
On this branch, holomorphy, symmetries, and limits fix the exact
superpotential written in terms of invariants, to be
\begin{eqnarray}\label{Eqn:Wexactfull}
W & = &  ~\lambda \Phi ~{\rm Tr} M +
(N_c - N_f) \left[ { \Lambda^{3N_c-N_f} \over
   {\rm det}~M } \right]^{1/(N_c-N_f)}
  \\ \nonumber 
& + & 
\lambda^{\prime} \Phi  ~{\rm Tr}  M^{\prime}  +
(N_c^{\prime} - N_f^{\prime})
  \left[ { \Lambda^{\prime 3N_c^{\prime}-N_f^{\prime}} \over
   {\rm det}~M^{\prime} }
   \right]^{1/(N_c^{\prime}-N_f^{\prime})}\,.
\end{eqnarray}
For gauge sectors in the free magnetic range, the nonperturbative contribution
refers to the Seiberg dual.
Since the meson invariants are lifted on this branch, they may
be eliminated by equations of motion
to give the exact superpotential in terms of the classical
modulus $\Phi$
\begin{eqnarray}\label{Eqn:Wexactgaugino}
W &=& N_c \left[ (\lambda \Phi)^{N_f}  \Lambda^{3N_c-N_f}
    \right]^{1/N_c}\nonumber\\
& & + N_c^{\prime} \left[ (\lambda^{\prime} \Phi)^{N_f^{\prime}}
   \Lambda^{\prime 3N_c^{\prime}-N_f^{\prime}}
    \right]^{1/N_c^{\prime}}\,.
\end{eqnarray}
The supersymmetric minima are given by stationary
points of the superpotential, $\partial W/ \partial \Phi =0$.

The metastable vacua found below occur when one of the 
SQCD sectors (the unprimed sector) is in the free magnetic range, i.e.
$N_c+1 \le N_f < {3 \over 2} N_c$, and the other (primed) sector exhibits an ADS runaway 
behavior, i.e. $N_f'<N_c'$ \cite{Affleck:1983mk}.
Moreover, we work in the range $\Lambda'$ $\ll$ $E$ $\ll$ $\Lambda$, where 
the primed sector is weakly interacting while the unprimed sector
has a dual weakly coupled description in terms of the magnetic
gauge group $SU(\tilde N_c)$ with $\tilde N_c$=$N_f-N_c$, $N_f^2$ singlets $M_{ij}$, and $N_f$
magnetic quarks $(q_i,\, \tilde q_j)$. In terms of this description, the full nonperturbative
superpotential reads
\begin{eqnarray}\label{Eqn:Wclnp}
W &=& ~m\Phi~ {\rm tr}~M + h~{\rm tr}~ q M \tilde q +  \lambda'\Phi~ {\rm tr}~ P \bar P \nonumber\\
& & +(N'_c - N'_f) \left( { \Lambda'^{3N'_c - N'_f}} \over {{\rm det}\, P \bar P} \right) ^{1/(N'_c - N'_f)} \nonumber\\
& & +(N_f-N_c)\left({{\rm det} \,M \over \tilde \Lambda^{3N_c - 2N_f}}\right)^{1/(N_f-N_c)}.
\end{eqnarray}
Hereafter, $M_{ij} = Q_{i} \bar Q_{j}/\Lambda$, and  $m:=\lambda \Lambda$.
The magnetic sector has a Landau pole at $\tilde{\Lambda}$ = $\Lambda$.

In this description, the meson $M$ and the primed quarks $(P, \,\bar{P})$ become massless at $\Phi=0$.
$M=0$ is also an enhanced symmetry point since here the magnetic quarks $(q, \, \tilde{q})$ become massless.

\section{Metastability near enhanced symmetry points}\label{sec:2}

Starting from the superpotential (\ref{Eqn:Wclnp}), the discussion is simplified by taking the limit
$\tilde \Lambda$ $\to$ $\infty$, while keeping $m$ fixed.
The nonperturbative ${\rm det}\, M$ term is only relevant for generating supersymmetric vacua, and not important for the details of the metastable vacua that
will arise near $M=0$.
Thus, for $M/\tilde \Lambda$ $\rightarrow$ $0$ and $\Phi /\tilde \Lambda$ $\rightarrow$ $0$, it
is enough to consider the superpotential
\begin{eqnarray}\label{Eqn:Wfullmagn}
W & = & m\Phi~ {\rm tr}~M + h~{\rm tr}~q M \tilde q + \lambda'\Phi \,{\rm tr}\, P \bar P \nonumber \\
& & + (N'_c - N'_f) \left( { \Lambda'^{3N'_c - N'_f}} \over {{\rm det}\,P \bar P} \right) ^{1/(N'_c - N'_f)}\,.
\end{eqnarray}

As discussed in~\cite{Essig:2007xk}, metastable vacua can only appear in
the region near the enhanced symmetry point $M=0$.  This still has a runaway, but
it turns out that one-loop corrections stop it; this novel effect is characterized as
a ``pseudo-runaway''.
The reason for this is that the Coleman-Weinberg formula
\beq\label{Eqn:Vcworiginal}
V_{CW}={1 \over {64 \pi^2}}\,{\rm Str} \,  M^4~ {\rm ln}~ M^2
\eeq
will have polynomial (instead of logarithmic) dependence. This will be explained next.

In the region $\Phi \neq 0$, $(P, \bar P)$ may be integrated out by equations of motion provided that $\Lambda' \ll \lambda' \Phi$. This is a good
description if we are not exactly at the origin but near it, as given by $\Phi / \tilde \Lambda \ll 1$. Taking, as before,
$\tilde{\Lambda} \to \infty$ and $m$ fixed, the superpotential reads
\begin{eqnarray}\label{Eqn:Wdimple}
W & = & m\Phi\, {\rm tr}\,M+h \,{\rm tr}\,q M \tilde q \nonumber \\
& & + N_c' \big[\lambda'^{N_f'} \,\Lambda'^{3N_c'-N_f'}\,\Phi^{N_f'}\big]^{1/N_c'}\,.
\end{eqnarray}
This description corresponds to
an O`Raifeartaigh - type model in terms of magnetic variables but with no
flat directions.

Given that $\phi=\langle \Phi \rangle \neq 0$, we will expand around the point of maximal symmetry
\begin{eqnarray}\label{Eqn:issvac}
& & q=\pmatrix{q_0&0},\,\tilde q=\pmatrix{\tilde q_0 \cr 0},\, \nonumber \\
& & M=\pmatrix{0&0\cr0& 0+X \cdot I_{N_c\times N_c}}.
\end{eqnarray}
Here $q_0$ and $\tilde q_0$ are $\tilde N_c \times \tilde N_c$ matrices satisfying
\beq\label{Eqn:qo}
h q_{0i} \tilde q_{0j}=-m\phi\,\delta_{ij}\;,\;i,j=\tilde N_c+1,\ldots N_f\,,
\eeq
and the nonzero block matrix in $M$ has been taken to be proportional to the
identity; indeed, only ${\rm tr}\,M$ appears in the potential. This minimizes $W_M$ and
sets $W_q=W_{\tilde q}=0$. The spectrum of fluctuations around (\ref{Eqn:issvac}) 
was studied in detail in \cite{Essig:2007xk},
where it was shown that the lightest degrees of freedom correspond to \\
$(\phi,  X)$ with mass given by $m$. The effective potential derived from (\ref{Eqn:Wdimple}) is
\begin{eqnarray}\label{Eqn:Vdimple}
V(\phi, X) & = & \Bigg| m N_c X+N_f' \lambda'^{N_f'/N_c'}
\left({\Lambda'^{3N_c'-N_f'} \over \phi^{N_c'-N_f'}}\right)^{1/N_c'}\Bigg|^2 \nonumber \\ 
& & + N_c m^2 |\phi|^2 + V_{CW}(\phi, X)\,,
\end{eqnarray}
where the first term comes from $W_\phi$.
The Coleman-Weinberg contribution will be discussed shortly.

As a starting point, set $ X=0$ and $V_{CW} \to 0$. Minimizing $V(\phi,  X=0)$ gives
\beq\label{Eqn:pdimple}
|\phi_0|^{(2N_c'-N_f')/N_c'}=\sqrt{{N_c'-N_f'} \over {N_c N_c'}}\,\,N_f' {\lambda'^{N_f'/N_c'} \over m} \Lambda'^{\frac{3N_c'-N_f'}{N_c'}}
\eeq
and since $W_{\phi \phi} \sim m$, $V(\phi_0+\delta \phi, X=0)$ corresponds to a parabola of
curvature $m$. The nonperturbative term only affects $\phi_0$ but not the curvature $m$.

Next, allowing $X$ to fluctuate (but still keeping $V_{CW} \to 0$), $V(\phi_0, X)$
gives a parabola centered at
\beq\label{Eqn:Xdrain}
X_{W_{\phi}=0} = - {\sqrt{N_c'\over{N_c(N_c'-N_f')}}}|\phi_0|
\eeq
and curvature $m$. In other words, $X=0$ is on the side of a hill of curvature $m$ and height
$V(\phi_0, 0) \sim m^2 |\phi_0|^2$.

To create a minimum near $X=0$, $V_{CW}$ should contain a term $m_{CW}^2 |X|^2$,
with $m_{CW}$ $\gg$ $m$; this would overwhelm the classical curvature. As explained in \cite{Essig:2007xk}, the massive degrees of freedom giving the dominant
contribution
to $V_{CW}$ come from integrating out the massive fluctuations along $q_0$ and $\tilde q_0$. The result is
\beq\label{Eqn:Vcw}
V_{CW}=N_c bh^3 m |\phi| |X|^2+\ldots
\eeq
with $b$=$({\rm log}4-1)/8\pi^2 \tilde N_c$ \cite{Intriligator:2006dd}, and `$\ldots$' represent 
contributions that are unimportant for the present discussion. 
In this computation, $X$ and $\phi$ are taken as background fields. 
The quadratic $X$ dependence appears because $X=0$ is an enhanced symmetry point.

In order to be able to produce a local minimum, the marginal parameters $(\lambda ,\lambda')$ will have to be tuned to satisfy
\beq\label{Eqn:branch}
\epsilon \equiv {{m^2} \over m^2_{CW}} = {m\over bh^3 |\phi|} \ll 1 \,.
\eeq
In this approximation, the value of $\phi$ at the minimum is still given by (\ref{Eqn:pdimple}); also, $X$
is stabilized at the nonzero value
\beq\label{Eqn:Xdimple}
|X_0|=\sqrt{N_c N_c' \over{N_c'-N_f'}}\,{m \over {bh^3}}\,,
\eeq
with $|X_0|\ll X_{W_{\phi}=0}$.  
At the minimum, (\ref{Eqn:branch}) gives
\beq\label{Eqn:brancha}
(m/\Lambda')^{3N_c'-N_f'}\ll (bh^3)^{(2N_c'-N_f')/N_c'}\lambda'^{N_f'}
\eeq
so the Yukawa coupling $\lambda$ in $m=\lambda \Lambda$ must be taken small for the analysis to
be self-consistent.
The calculability condition $\Lambda' \ll \lambda' \Phi$ follows as a consequence of this.
At the minimum, $X_0 \ll \phi_0$.  The F-terms are given by $W_\phi$ $\sim$ $m\phi_0$ $\sim$ $W_X$,
and from (\ref{Eqn:pdimple}) the scale of supersymmetry breaking is thus controlled
by the dynamical scales of both gauge sectors.

\begin{figure}
\includegraphics[scale=0.24]{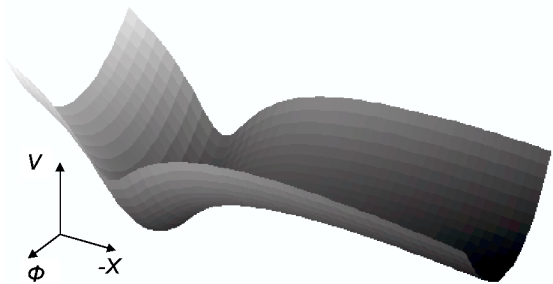}
\caption{A plot showing the shape of the potential, including the
one-loop Coleman-Weinberg corrections, near the metastable minimum.
In the $\phi$-direction the potential is a parabola, whereas in the $X$-direction
it is a side of a hill with a minimum created due to quantum corrections.
This plot was generated with the help of \cite{vandenBroek:2007kj}.}
\label{fig:1}
\end{figure}

Thus the model has a metastable vacuum near the origin, created by a combination of quantum corrections and
nonperturbative gauge effects.  The pseudo-runaway towards $X=X_{W_{\phi}=0}$ has been lifted by the
Coleman-Weinberg contribution, as anticipated.  This is the origin of the $1/b$ dependence in (\ref{Eqn:Xdimple}).
The local minimum is depicted in Fig.~\ref{fig:1}. 

The metastable vacuum appears from a competing effect between a runaway behavior
in the primed sector and one loop corrections for the meson field $X$. One is naturally led
to ask if, under these circumstances, other quantum effects are under control. These include higher loop
terms from the massive particles producing $V_{CW}$ as well as perturbative $g'$ corrections. In~\cite{Essig:2007xk} it was shown that such corrections are under control and do not destabilize the metastable vacuum.

\begin{figure}
\includegraphics[scale=0.4]{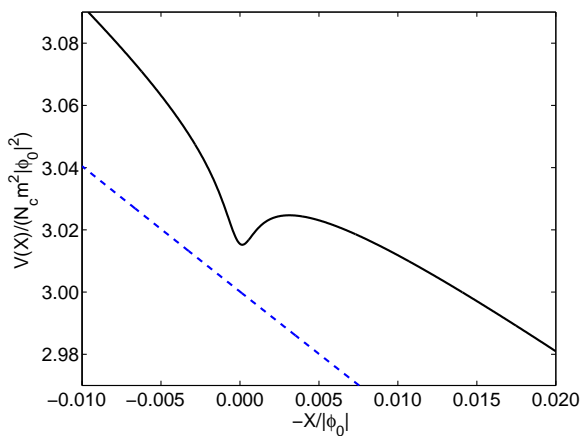}
\caption{The classical potential (dashed line) and the total potential
including one-loop corrections (solid line) for fixed $|\phi|=|\phi_0|$,
where $|\phi_0|$ is the value of $\phi$ at the metastable minimum (\ref{Eqn:pdimple}).
In the figure, $N_f=3$, $N_c=2$, $N_f'=1$ and $N_c'=2$, and 
$W_{\phi}=0$ has been scaled to equal 1 on both axes;
the metastable minimum lies at $\sim 10^{-4}$.
This plot was generated with the help of \cite{vandenBroek:2007kj}.}
\label{fig:2}
\end{figure}

The metastable non-supersymmetric vacuum can be made parametrically long-lived by taking the parameter
$\epsilon \equiv {m \over {b h^3 |\phi_0|}}$ sufficiently small.
The direction in field space to tunnel out of the false vacuum
is towards negative $X$ with fixed $|\phi|=|\phi_0|$ \cite{Essig:2007xk};
the one-dimensional potential $V(X)$ $\equiv$ $V$($|\phi|$=$|\phi_0|$,$X$) is shown in Fig.~\ref{fig:2}.
The bounce action scales as
\beq\label{Eqn:bounce}
B \sim {{\tilde{X}^4}\over{V(X_{peak})-V(X_{0})}}
  \sim b\, h^3 \, {1\over \epsilon^2},
\eeq
where the field tunnels to $\tilde{X}$, and $V(X)$ reaches a local maximum at 
$X_{peak}$.  Thus $B\rightarrow\infty$ as $\epsilon\rightarrow 0$.
 
To have gaugino masses, any R-symmetry must be broken, explicitly and/or spontaneously. 
The low energy superpotential (\ref{Eqn:Wdimple}) has the $U(1)_R$ symmetry
\beq\label{Eqn:Rdimple}
R_{\phi}=2 {N_c' \over N_f'}\;,\;R_X=2 {{N_f'-N_c'} \over N_f'}\;,\;R_q=R_{\tilde q}={N_c' \over N_f'}\,.
\eeq
Since the VEV's of these fields are nonzero in the metastable vacuum, the R-symmetry is spontaneously broken, and there is
an R-axion $a$. 
For finite $\tilde \Lambda$, the ${\rm det}~X$ contributions need to be taken into account, and the $U(1)_R$ symmetry
becomes anomalous. 
This explicit breaking gives a mass to the R-axion,
\beq\label{Eqn:axionm}
m_a^2 \sim m^2\,\Bigg(\bigg[{\lambda \over {bh^3}}\bigg]^{2\,{{3N_c-2N_f}\over{N_f-N_c}}}\,{\epsilon \over{bh^3}} \Bigg)\ll m^2\,,
\eeq
where $\lambda$ is the Yukawa coupling appearing in $m=\lambda \Lambda$.

\section{Meta-Stability Near Generic Points of Enhanced Symmetry}\label{sec:3}

A generic structure in the landscape of effective field theories corresponds to a gauge theory with vector-like matter
and mass given by a singlet, whose dynamics is related to another sector. The superpotential may be written as
\beq\label{Eqn:genw}
W=f(\Phi)+\lambda\,\Phi {\rm tr}(Q \bar Q)\,.
\eeq
Here, $(Q, \bar Q)$ are $N_f$ quarks in $SU(N_c)$ SQCD; $f(\Phi)$ may be generated, for instance, from a flux superpotential, by
nonrenormalizable interactions, or by another gauge sector. Next, it is required that the SQCD sector be in the free magnetic range, 
which is still a generic situation. The dual magnetic description is weakly coupled near the enhanced symmetry point $\Phi=0$, 
where the superpotential reads
\beq\label{Eqn:genw2}
W=f(\Phi)+m \Phi\, {\rm tr}~M + h\, {\rm tr}~q M \tilde q\,.
\eeq

The question that will be addressed here is: what restrictions need to be imposed on $f(\Phi)$, so that the one loop potential
$V_{CW}$ can create a metastable vacuum near $M=0$?  Since we are interested in the novel effect of pseudo-runaway
directions we will demand $f'(\Phi)\neq 0$.  The model in \cite{Brummer:2007ns} would correspond to $f'(\Phi)=0$.

As discussed in Sec.~3, this is possible only if
\beq\label{Eqn:restrm}
m_{CW}^2:=N_c bh^3 m |\phi| \gg m^2\,
\eeq
where $\phi$ denotes the expectation value of $\Phi$ at the metastable vacuum. Further, one needs to impose that
\beq\label{Eqn:restrx}
h^2 |X|^2 \ll m |\phi|
\eeq
in order for the Taylor expansion of $V_{CW}$ around $X=0$ to converge. Evaluating the potential as in (\ref{Eqn:Vdimple}),
\beq\label{Eqn:vgen}
V=N_c m^2 |\phi|^2+ \big|f'(\phi)+m N_c\, X \big| ^2+m_{CW}^2 |X|^2\,.
\eeq
The rank condition, an essential ingredient in the discussion, just follows from having SQCD in the free magnetic range.
This fixes the first term, which comes from $W_M$, and the block structure of the matrix $M$; $X$ was defined in (\ref{Eqn:issvac}).

The minima of $V(\phi, X)$ are 
\beq\label{Eqn:phiextr}
N_c m^2 \phi=-f'(\phi)\, f''(\phi)^*,\;m_{CW}^2 \, X=-N_c m f'(\phi)\,.
\eeq
It is possible to combine the conditions (\ref{Eqn:restrm}) and (\ref{Eqn:restrx})
with the values at the metastable vacuum (\ref{Eqn:phiextr}) to derive constraints on $f(\phi)$: (\ref{Eqn:restrm}) now reads
\beq\label{Eqn:restrma}
\frac{|f'(\phi)f''(\phi)|}{m^3} \gg {1 \over {bh^3}}\,,
\eeq
while (\ref{Eqn:restrx}) gives
\beq\label{Eqn:restrxa}
h^2 |f'(\phi)|^2 \ll m (bh^3)^2 |\phi|^3\,.
\eeq
The necessary conditions for metastable vacua near $X=0$ to exist are (\ref{Eqn:restrma}) and (\ref{Eqn:restrxa}). 
These conditions may require fine-tuning the coefficients of $f(\phi)$, for example if one has two gauge sectors as 
in (\ref{Eqn:Wtreem}) with non-coincident enhanced symmetry points \cite{Essig:2007xk}. 
In the case of coincident enhanced symmetry points, where there are no relevant scales, no fine-tuning is required.

\vspace{0.3cm}
In summary, we constructed a model with long-lived metastable vacua in which
all the relevant parameters, including the supersymmetry breaking scale, are generated dynamically
by dimensional transmutation.
The model has the desirable features of an explicitly and spontaneously broken
$R$-symmetry, a singlet, a large global symmetry, naturalness and renormalizability.
The metastable vacua are produced near a point of enhanced symmetry by a combination
of nonperturbative gauge effects and, crucially, perturbative effects coming from the one-loop
Coleman-Weinberg potential.
A salient feature is the existence of pseudo-runaway directions, which correspond to a runaway 
behavior that is lifted by one loop quantum corrections.
It would be interesting to study the implications of this for the landscape of supersymmetric gauge 
theories.

\vspace{0.31cm}
\noindent{\emph{Acknowledgments:}}

We thank S.~Thomas for suggesting this problem.  
RE would like to thank the organizers of SUSY07 for the opportunity 
to present this work.
This research is supported by the Department of Physics and Astronomy 
at Rutgers University.

%
%

\end{document}